\documentclass[%
 preprint, 
 amsmath,amssymb,
 aps, physrev,
]{revtex4-2}

\usepackage{graphicx}
\usepackage{dcolumn}
\usepackage{bm}
\usepackage{float}
\usepackage[labelfont=bf,textfont=bf]{caption}
\usepackage{natbib}

\begin{document}


\title{\textbf{10-W Sub-100-fs Ultrafast Cr:ZnS/ZnSe MOPA System enabled by doping gradient engineering}}

\author{\textbf{Guangzi Feng}$^{1,2,\dagger}$}
\author{\textbf{Xiyue Zhang}$^{1,3,\dagger}$}
\author{\textbf{Yuchen Wang}$^{1,\ddagger}$}
\author{\textbf{Weibo Wu}$^{4}$}
\author{\textbf{Gianluca Galzerano}$^{4,5}$}
\author{\textbf{Qing Wang}$^{6}$}
\author{\textbf{Ting Yu}$^{1}$}
\author{\textbf{Yujie Peng}$^{1,7}$}
\author{\textbf{Jintai Fan}$^{1}$}
\author{\textbf{Benxue Jiang}$^{8,\ddagger}$}
\author{\textbf{Yuxin Leng}$^{1,7,\ddagger}$}
\author{\textbf{Long Zhang}$^{1}$}

\affiliation{$^{1}$Shanghai Institute of Optics and Fine Mechanics, Chinese Academy of Sciences, Shanghai 201800, China}
\affiliation{$^{2}$ShanghaiTech University, Shanghai 200120, China}
\affiliation{$^{3}$University of Chinese Academy of Sciences, Beijing 100049, China}
\affiliation{$^{4}$Politecnico di Milano, Milano 20133, Italy}
\affiliation{$^{5}$IFN-CNR, Milano 20133, Italy}
\affiliation{$^{6}$School of Optics and Photonics, Beijing Institute of Technology, Beijing 100081, China}
\affiliation{$^{7}$State Key Laboratory of Ultra-intense laser Science and Technology, Shanghai Institute of Optics and Fine Mechanics, Chinese Academy of Sciences, Shanghai 201800, China}
\affiliation{$^{8}$School of Physics Science and Engineering, Tongji University, Shanghai 200092, China}

\thanks{$\dagger$ These authors contributed equally to this work.\\
$\ddagger$ Corresponding authors: wangyuchen@siom.ac.cn; 24070@tongji.edu.cn; lengyuxin@siom.ac.cn}



\raggedbottom             
\begin{abstract}

We report on a high-power mid-infrared femtosecond master oscillator power amplifier (MOPA) system, employing Cr:ZnS and Cr:ZnSe polycrystals with fine-tuned doping profiles. Based on the soft-aperture Kerr-lens mode-locking in the soliton regime, the seed oscillator generates $\sim$40-fs pulses with a repetition rate $\sim$173 MHz with an average power close to 400 mW. The amplification process of the seed pulse train is investigated in depth in a single-pass configuration for both Cr:ZnS and Cr:ZnSe crystal rods. For further power scaling, a dual-stage MOPA system has been implemented, generating pulse trains with an average power up to 10.4 W, limited only by the pump source, with a re-compressed pulse duration of 78 fs using a dispersion compensator comprising chirped mirrors and sapphire plates. This work paves the way for further power scaling of mid-infrared Cr:ZnS/ZnSe ultrafast laser systems without moving parts for applications in material processing, remote sensing and medicine.
\begin{description}

\item[Key words]
High power lasers; Mid-infrared (MIR) lasers; Ultrafast lasers; Cr:ZnS/ZnSe crystals.
\end{description}
\end{abstract}

\maketitle


\section{Introduction}

The mid-infrared (mid-IR) spectral region (generally considered to be from 2.5 to 25 {\textmu}m) has received increasing attention in recent years, in part for the presence of the so-called `molecular fingerprint' region \cite{Diddams2007,Huber2024,Schliesser2012}. With the ground-breaking invention of optical frequency combs and their spectroscopy techniques, direct real-time access to the fundamental ro-vibrational transitions of gaseous molecules are within reach, at levels of accuracy and precision never imagined before\cite{Picque2019,Coddington2016}. With recent advances in the field of AMO (atomic, molecular and optical) physics and precision spectroscopy, mid-IR transitions are being probed for molecular hydrogen ions \cite{Alighanbari2025}, muonic hydrogen \cite{Pohl2010} and trichloromethane\cite{Gambetta2017}, which are expected to bring new insights ranging from the constancy of proton-electron mass ratio to the revision of the proton radius. 

As the centerpiece of a frequency comb system, robust and potent ultrafast lasers have long been in need for this particular spectral region, to reduce the complexity and bulkiness of current sources based on nonlinear optical methods including difference frequency generation and optical parametric oscillators \cite{Schliesser2012,Hoghooghi2022,Ruehl2012,Liendecker2011}. To this end, transition metal (TM)-doped II–VI semiconductors have been envisioned as a potential platform for mid-IR high-power and ultrafast lasers covering the 2--5~{\textmu}m range \cite{Mirov2018,Ma2019,Pupeza2015}. Among them, Cr\textsuperscript{2+}:ZnS/ZnSe show wide emission spectra, superior thermo-optical and mechanical stabilities, therefore being considered the ideal platform for ultrafast laser systems in the 2--3~{\textmu}m region, and often referred to as the `Ti\textsuperscript{2+}:sapphire of the mid-IR' \cite{Sorokina2003}. With a continuous development in the past 20 years, ultrafast Cr\textsuperscript{2+}:ZnS/ZnSe laser oscillators has seen tremendous progress, from active mode-locking\cite{Carrig2000}, passive mode-locking with saturable absorbers\cite{Cizmeciyan2013,Okazaki2019}, passive mode-locking with Kerr-lens effect (KLM)\cite{Tolstik2013,Wang2017,Nagl2019,Vasilyev2021,Wang2022} to mode-locking at GHz-repetition rates\cite{Barh2022}. These developments have cherished in the demonstration of a frequency comb based on Cr:ZnS \cite{Vasilyev2019} and the coherent spectral extension in single-cycle regimes \cite{Steinleitner2022,Kowalczyk2023}. The power scaling of the Cr\textsuperscript{2+}:ZnS/ZnSe continuous-wave (CW) and ultrafast lasers has been thoroughly investigated at University of Alabama at Birmingham, lead by S. Mirov and V. Gapontsev, with notable results in an 140-W CW Cr:ZnSe MOPA system \cite{Moskalev2016} and a 27-W fs MOPA demonstration \cite{Vasilyev2018}. These demonstrations rely on an innovative rotating gain element to keep the thermal effects at bay. Without any moving parts, the highest output power of Cr:ZnS/ZnSe fs MOPAs is still limited to $\sim$7.4 W\cite{Vasilyev2017,lu2023terawatt,HU2025112924}. To further tackle the limitations posed by thermal-optical effects it is necessary to revisit the gain element itself, with a proposal to exploit the inherent concentration gradient in the thermally diffused Cr:ZnS/ZnSe crystals \cite{kurashkin2018doping,Zhang2025}.

Here, we report on the power scaling with a master oscillator power amplifier (MOPA) system based on engineered doping profiles of the Cr:ZnS/ZnSe crystals, boosting the pulse train from a Cr:ZnS ultrafast oscillator to an average power beyond 10~W, without the use of any moving parts for cooling. A 27-mm-long Cr:ZnS polycrystal rod and a 28-mm-long Cr:ZnSe are fabricated for the amplifier, with central concentrations being $0.5 \times 10^{18}$~cm$^{-3}$ and $1.07 \times 10^{19}$~cm$^{-3}$ respectively, and concentrations ranging from $1.84 \times 10^{19}$~cm$^{-3}$ to $8.07 \times 10^{20}$~cm$^{-3}$. A Kerr-lens mode-locked Cr:ZnS resonator is employed as the master oscillator, which features an output pulse duration of 39~fs and an average power of 350~mW at a repetition rate of 173~MHz, corresponding to a pulse energy of 2~nJ. In terms of amplifiers, three different architectures have been investigated, including a single-pass Cr:ZnS amplifier displaying a maximum output power of 2.35~W; a single-pass Cr:ZnSe amplifier, which exhibited a maximum output power of 4.31~W and a small-signal gain of 38 times; and the dual-stage MOPA system demonstrating a maximum output power of 10.4~W, achieving a 30-fold amplification of the seed pulses, which presents a corresponding pulse energy of 60.1~nJ with barely any gain narrowing observed in the output spectrum. The pulse duration after dispersion compensation is measured to be 78~fs.

\section{Fabrication of Cr:ZnS/ZnSe polycrystal rods}

\begin{figure}[H]
    \centering
    \includegraphics[width=0.9\linewidth]{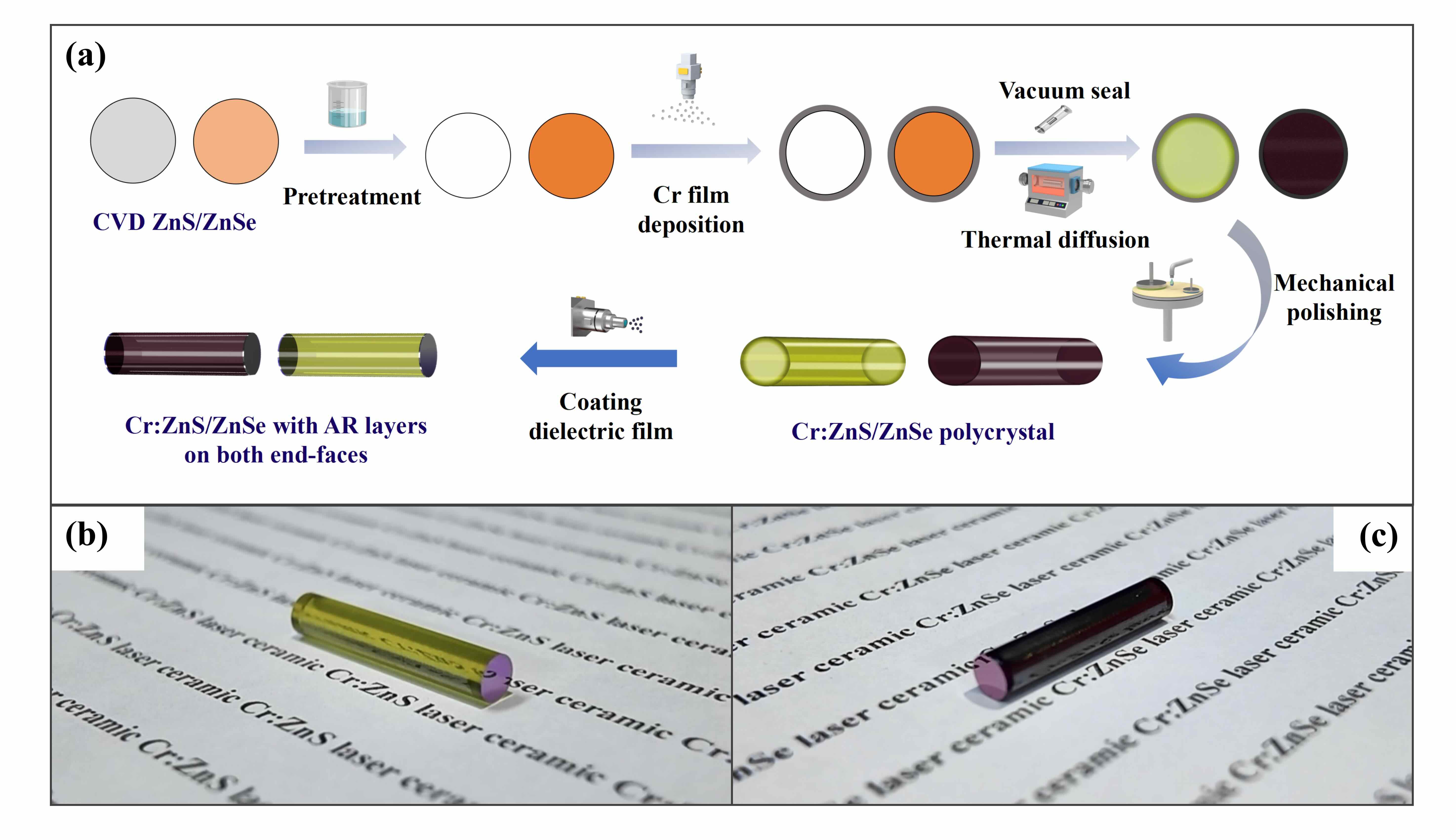}
    \caption{(a) The process flow for the preparation of the Cr:ZnS/ZnSe polycrystal rods. 
    (b) Appearance of the Cr:ZnS rod. 
    (c) Appearance of the Cr:ZnSe rod.}
    \label{fig:fabrication}
\end{figure}

The fabrication of Cr\textsuperscript{2+}-doped ZnS/ZnSe polycrystals involved the following sequential process, as shown in Fig.~\ref{fig:fabrication}. ZnS substrates and ZnSe substrates (Sinoma Artificial Crystal Research Institute)  were synthesized via chemical vapor deposition (CVD).  Surface pretreatment was included mechanical polishing and ultrasonic cleaning to eliminate particulate contaminants. The substrates were subsequently immersed in acetone for 5 minutes to dissolve organic residues. To enhance atomic adhesion, they were etched in anhydrous nitric acid solution for 30 minutes to form fresh micro-roughened surfaces, followed by a final rinse with ethanol. A chromium thin film approximately 1 {\textmu}m thick was deposited on the prepared substrates using filtered arc deposition (FAD). The chromium-coated samples were sealed in vacuum-tight quartz ampoules and subjected to thermal diffusion in a horizontal tube furnace at 960~$\pm$~5 {\textcelsius} for 720 hours (30 days) tofacilitate ion interdiffusion, resulting in Cr\textsuperscript{2+}:ZnS/ZnSe rod-shaped polycrystals. Finally, a dielectric antireflection (AR) coating for 1.5–3.0 {\textmu}m was deposited on the polished surfaces.

\begin{figure}[H]
    \centering
    \includegraphics[width=0.9\linewidth]{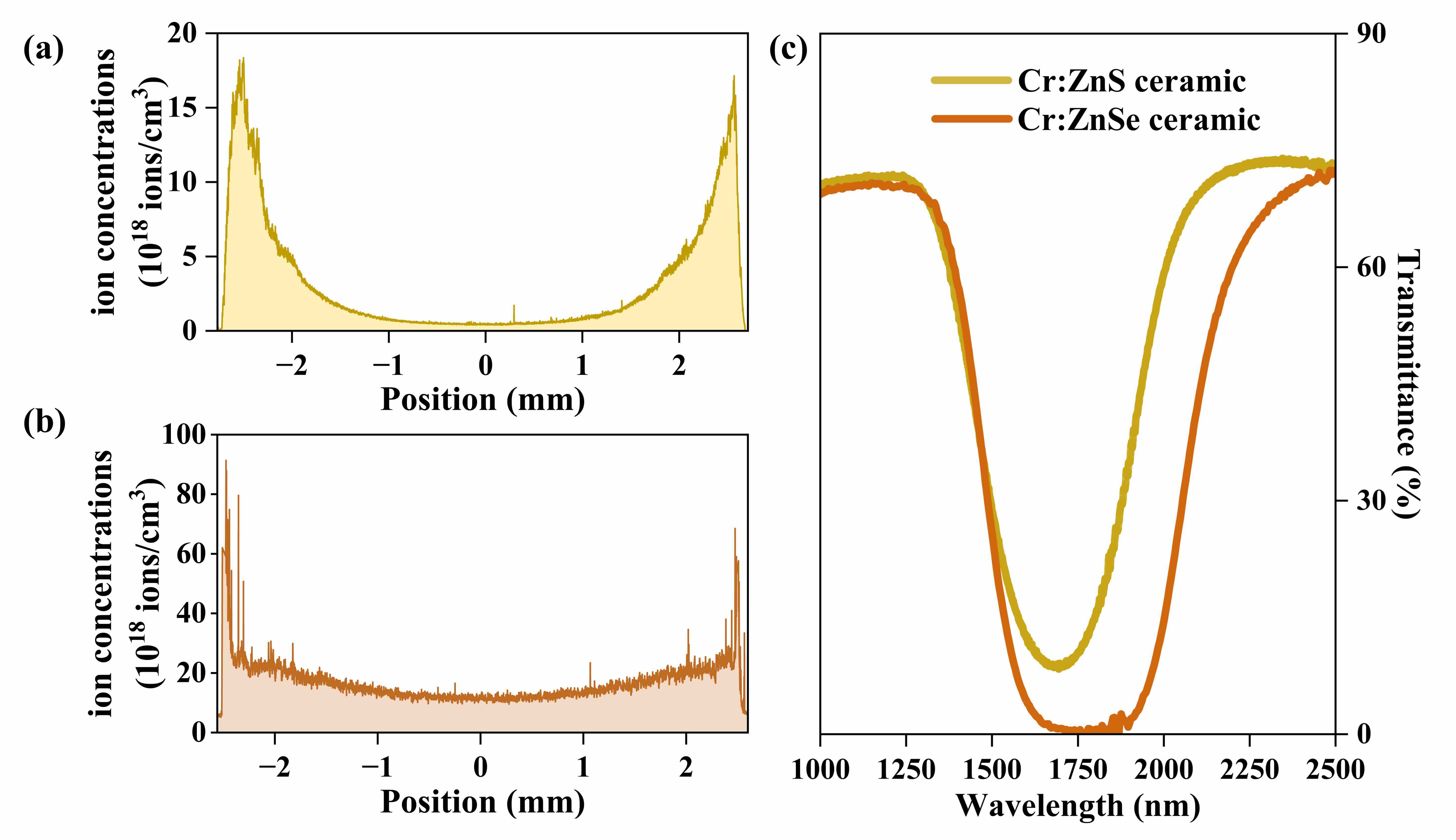}
    \caption{(a) Ion concentration profile of Cr:ZnS. 
    (b) Ion concentration profile of Cr:ZnSe. 
    (c) Transmittance spectra of Cr:ZnS and Cr:ZnSe polycrystals.}
    \label{fig:characterization}
\end{figure}

The spatial distribution of Cr\textsuperscript{2+} ions across the diameter of the end-faces of rod-shaped Cr:ZnS and Cr:ZnSe samples was analyzed via laser ablation inductively coupled plasma mass spectrometry (LA-ICP-MS). As depicted in Fig. \ref{fig:characterization}(a) and (b), the concentration profiles of Cr\textsuperscript{2+} reveal inhomogeneous distributions, with central concentrations of 0.5 $\times$ 10\textsuperscript{18} cm\textsuperscript{-3} in Cr:ZnS and 1.07 $\times$ 10\textsuperscript{19} cm\textsuperscript{-3} in Cr:ZnSe. Under same thermal diffusion conditions, the Cr:ZnSe demonstrates a substantially higher Cr\textsuperscript{2+} ion concentration with a reduced gradient, leading to a more homogeneous distribution.
Fig. \ref{fig:characterization}(c) and (d) display the transmittance spectra of Cr:ZnS and Cr:ZnSe polycrystals. Under examined doping conditions, the Cr:ZnS and Cr:ZnSe polycrystals show good transmission properties. The absorption spectra exhibit characteristic peaks at 1680 nm for Cr:ZnS and 1864 nm for Cr:ZnSe, attributed to the \textsuperscript{5}T\textsubscript{2}→\textsuperscript{5}E electronic transition of Cr\textsuperscript{2+} ions within their respective host lattices. Notably, both materials maintain high transmittance values at 2400 nm, with Cr:ZnS achieving 73.7\% and Cr:ZnSe 70.7\%, underscoring their suitability for mid-infrared applications.

\section{Kerr-Lens Mode-Locked Seed Oscillator}

\begin{figure}[H]
    \centering
    \includegraphics[width=0.95\linewidth]{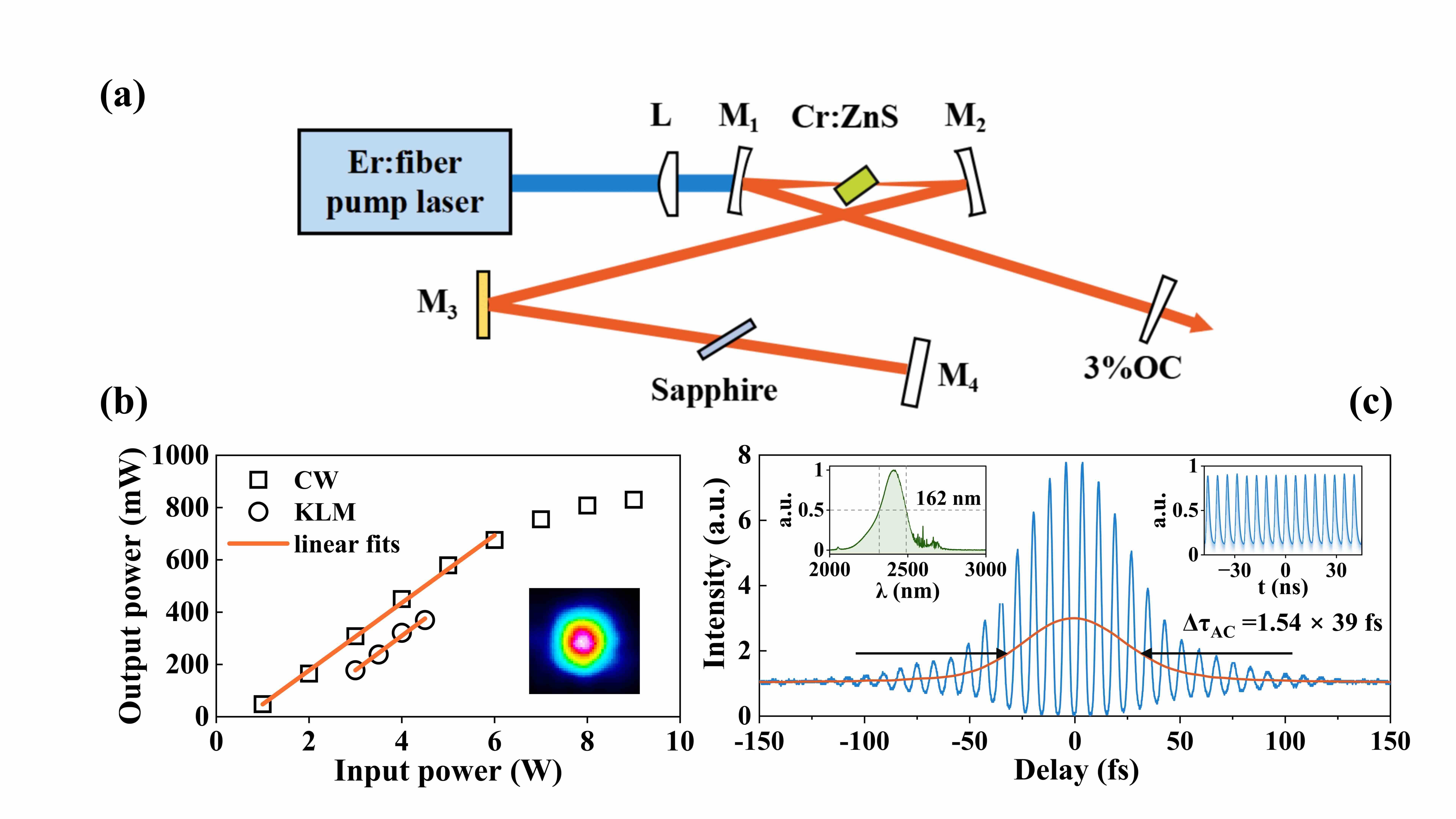}
    \caption{(a) Asymmetric linear cavity configuration of the KLM oscillator. 
    (b) Output power versus input power in both CW and KLM regimes. Inset: beam quality. 
    (c) Interferometric autocorrelation and spectrum (left inset), and pulse trains (right inset) of the KLM Cr:ZnS oscillator.}
    \label{fig:klm}
\end{figure}

As shown in Fig.~\ref{fig:klm}(a), a KLM cavity employing a 5-mm-long single-crystal Cr:ZnS (Miracrys) with a doping level of $1\times10^{19}$~ions/cm$^3$ as the gain medium is constructed with five dielectric mirrors: two plano-concave HR mirrors M$_1$ and M$_2$ (ROC = -100 mm), a plane chirped mirror M$_3$, a plane HR mirror M$_4$, and a wedged OC. The HR mirrors M$_1$, M$_2$, and M$_4$ have $R>99.5\%$ from 2000 to 2800~nm, $R<5\%$ from 1500 to 1650~nm, and $|$GDD$| < 100$~fs$^2$ from 2100 to 2500~nm (custom-made by Qifeng Landa Co. Ltd). M$_3$ has $R>99.5\%$ from 1700 to 2700~nm and a nominal GDD = -250~fs$^2$ (Ultrafast Innovations GmbH). The OC has $R=97\%$ from 2100 to 2400~nm (LAYERTEC GmbH). The total cavity length is 0.86~m, with the asymmetric HR arm to OC arm ratio of 3:5. The Cr:ZnS crystal, placed at the Brewster’s angle between M$_1$ and M$_2$, is mounted in a copper heat sink maintained at 20~{\textcelsius} by a Peltier cooler and temperature controller.

A single-mode, narrow-linewidth, linearly-polarized CW Er-doped fiber laser (CONNET, VFLS-1550-B-MP) with $\lambda=1550$~nm and maximum output power of 15~W is adopted as the pump source. A plano-convex lens ($f=100$~mm) focuses the pump beam into the Cr:ZnS crystal through M$_1$. The pump spot size is designed to be slightly smaller than the resonator fundamental mode to facilitate soft-aperture KLM, as calculated by the ABCD matrix method (ReZonator).

The pump power before M$_1$ and the output power from the OC were measured, as shown in Fig.~\ref{fig:klm}(b). In CW operation, the Cr:ZnS laser delivers a maximum output of 0.83~W at 9~W pump power. Self-starting KLM is achieved once the incident pump power exceeds 3~W with proper cavity alignment. At 7~W pump power, soliton mode-locking provides 572~mW output. The spot profile (Fig.~\ref{fig:klm}(b), inset) was measured at 50~cm from the OC using a mid-IR beam profiler (DataRay S-WCD-IR-BB-7.5).


The temporal profile is characterized with a home-built interferometric autocorrelator using two-photon absorption in an InGaAs detector (Thorlabs PDA20CS2), with residual chirp compensated by a 0.5-mm-thick Ge plate. The spectrum is measured with an FTIR spectrometer (Arcoptix FTIR-L1-120-4TE), while the pulse train and RF spectrum are measured with a fast HgCdTe photodetector ($\sim$1~GHz, Vigo Systems), oscilloscope (Tektronix MDO34), and ESA (Keysight N9020A).  

Autocorrelation traces at 173~MHz (Fig.~\ref{fig:klm}(c)) indicate $\sim$39~fs pulses, with TBP = 0.326, slightly above the transform-limited 0.315 (sech$^2$). The output spectrum is centered at 2411~nm (124.4~THz) with FWHM = 162~nm (9~THz); water vapor absorption beyond $\sim$2.5~{\textmu}m introduces spectral features, and the 10-dB bandwidth spans 559~nm. The right inset of Fig.~\ref{fig:klm}(c) shows stable soliton mode-locked pulse trains. The oscillator stays mode-locked for $>$2 hours without an enclosure.

\section{Single-Stage and Dual-Stage MOPA}

\subsection{Single-pass performance of Cr:ZnS polycrystal}

\begin{figure}[H]
    \centering
    \includegraphics[width=0.95\linewidth]{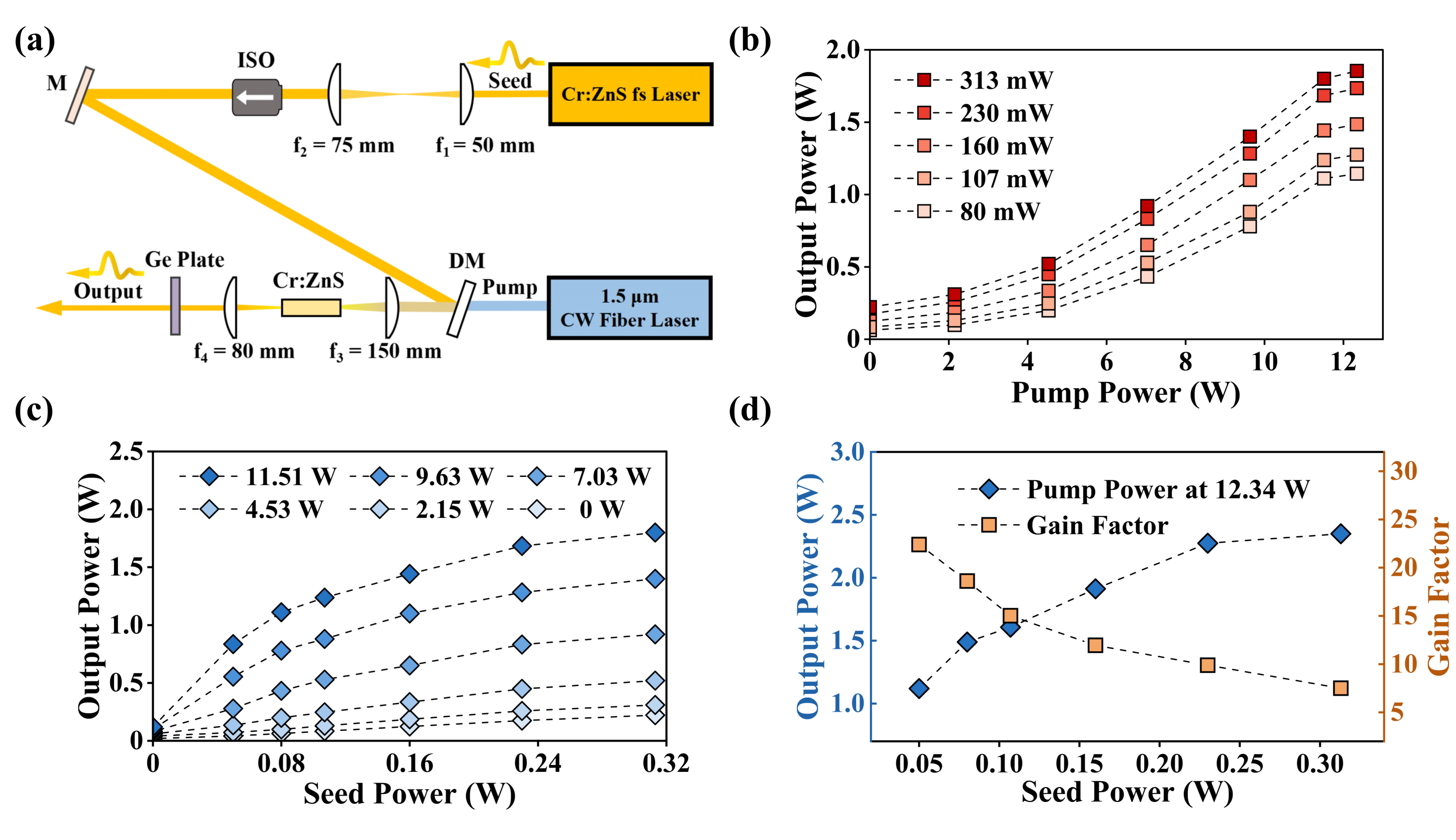}
    \caption{(a) Schematic of the Cr:ZnS MOPA system. 
    (b) Average output power versus CW pump power with different seed powers. 
    (c) Average output power versus seed power at 173~MHz with different pump powers. 
    (d) Output power at 12.34~W pump power (left y-axis) and gain factor (right y-axis) versus seed power.}
    \label{fig:mopa}
\end{figure}

A single-pass amplifier was constructed for the characterization of the Cr:ZnS rod, as shown in Fig.~\ref{fig:mopa}(a). The seed pulse emitted from the Cr:ZnS oscillator propagates leftward, first passing through two plano-convex lenses ($f_1=50$~mm, $f_2=75$~mm) forming a telescope for mode matching. An optical isolator downstream of $f_2$ ensures unidirectional transmission and suppresses back-reflection-induced instability. The collimated seed beam is then reflected by a gold mirror (M). Meanwhile, the collimated pump beam is combined with the seed via a dichroic mirror (DM), which transmits 1.55~{\textmu}m pump light and reflects the seed wavelength. The overlapped beams are focused into the Cr:ZnS crystal by a plano-convex lens ($f_3=150$~mm). The output beam is collimated by another lens ($f_4=80$~mm). After filtering out residual pump light using an AR-coated Ge window (2-6 {\textmu}m, Thorlabs), the output power is measured.  

The Cr:ZnS MOPA delivers a maximum pulsed output of 1.85~W under 313~mW seed injection (see Fig.~\ref{fig:mopa}(b)), at 12.3~W pump power, corresponding to 15\% optical-to-optical conversion efficiency. The output power exhibits a three-stage growth with increasing pump power: (i) at $<3$~W pump, output grows super-linearly as pump compensates intrinsic losses (absorption, scattering); (ii) in the linear regime ($3$–$10$~W), Cr:ZnS operates in pre-saturation amplification, gain remains stable, and higher seed powers trigger earlier linear scaling; (iii) at pump powers $>10$~W, slight deviation from linearity occurs due to combined gain saturation and thermal effects, where heat deposition in the crystal induces refractive index changes and mode distortion, reducing efficiency.  

Fig.~\ref{fig:mopa}(c) shows that output power increases monotonically with seed power across different pump powers, though the trend becomes sub-linear at higher seed powers. This deviation arises from gain saturation: as seed power rises, the finite inversion population—sustained by pump absorption—is depleted faster than replenished. Consequently, although output power continues to rise, the gain factor declines (Fig.~\ref{fig:mopa}(d)), indicating reduced amplification efficiency per unit seed power. This reflects the intrinsic limit of Cr:ZnS in maintaining population inversion under intense seed stimulation.  

\subsection{Single-pass performance of Cr:ZnSe polycrystal}

\begin{figure}[H]
    \centering
    \includegraphics[width=0.95\linewidth]{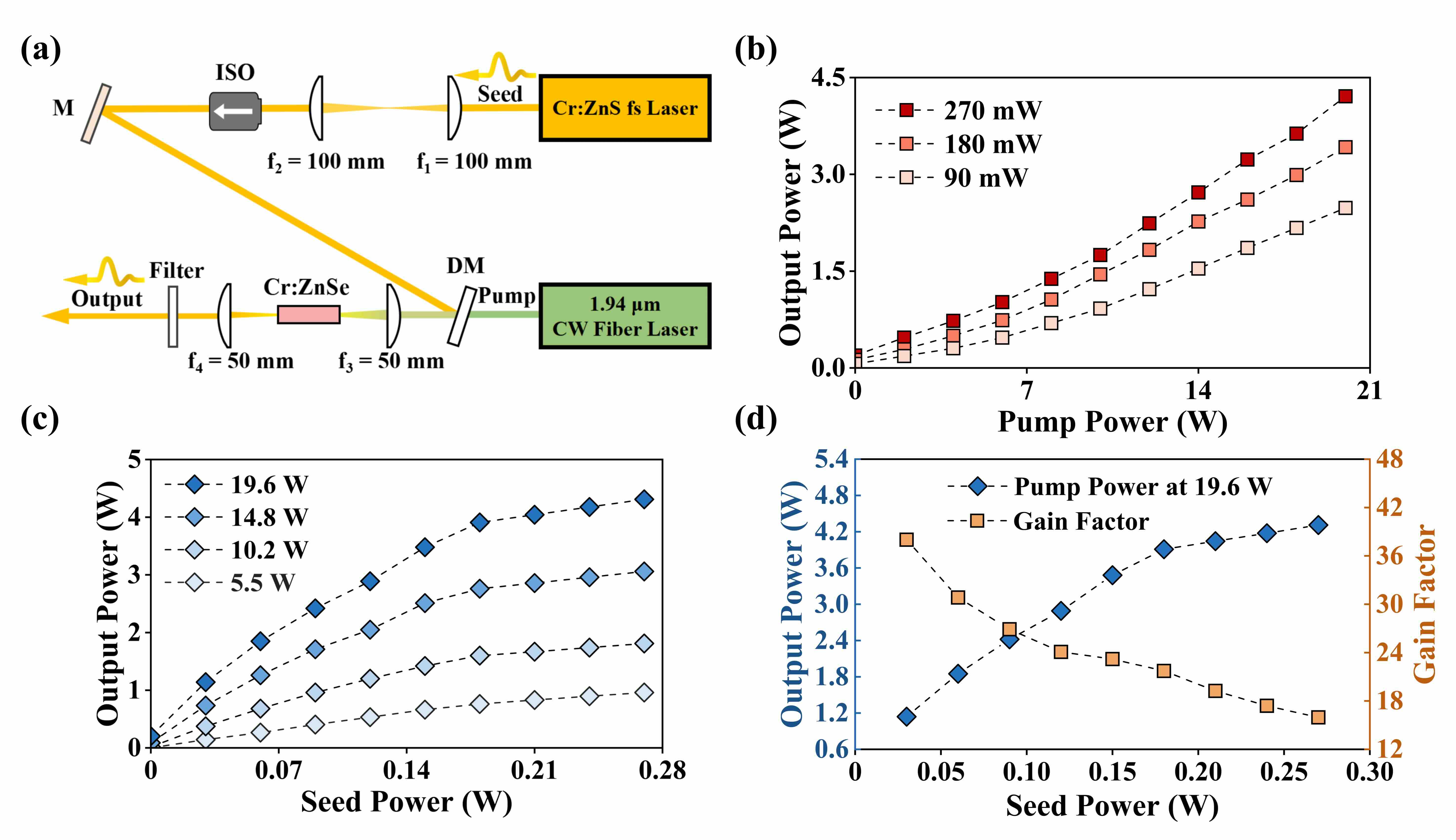}
    \caption{(a) Schematic of the Cr:ZnSe MOPA system. 
    (b) Average output power versus CW pump power with different seed powers. 
    (c) Average output power versus seed power at 173~MHz with different pump powers. 
    (d) Output power at 20~W pump power (left y-axis) and gain factor (right y-axis) versus seed power.}
    \label{fig:mopa_znse}
\end{figure}

For the Cr:ZnSe crystal rod, a similar single-pass amplifier is implemented, as shown in Fig.~\ref{fig:mopa_znse}(a), comprising a Cr:ZnS femtosecond seed laser, a CW 1.94~{\textmu}m Thulium-doped fiber pump laser (CONNET, 30-W), two pairs of plano-convex lenses ($f_1=f_2=100$~mm; $f_3=f_4=50$~mm), an optical isolator, a plane gold mirror, a dichroic mirror, and a 28-mm-long Cr:ZnSe crystal. The seed passes through the telescope ($f_2/f_1$) for beam-size adjustment, then through the isolator before reflection by M. The pump beam transmits through DM, which is highly reflective at the seed wavelength but highly transmissive at 1.94~{\textmu}m, ensuring spatial overlap in the Cr:ZnSe crystal. After passing the crystal, a longpass filter removes residual pump light, yielding spectrally pure 2.4~{\textmu}m output pulses.  

As seen in Fig.~\ref{fig:mopa_znse}(b), the Cr:ZnSe MOPA system amplifies 270~mW seed pulses (173~MHz repetition rate) to an average power of 4.21~W under 20~W pumping, achieving 21\% optical-to-optical efficiency. Unlike the sublinear trend in Cr:ZnS (Fig.~\ref{fig:mopa}(b)), the ZnSe amplifier exhibits linear scaling at low pump levels (0–5~W). No saturation was observed at 20~W, though pump power was limited to prevent possible thermal or optical damage.



The output power is measured to be increasing linearly with seed power, as shown in Fig.~\ref{fig:mopa_znse}(c), though slope efficiency declines slightly at high seed levels. As summarized in Fig.~\ref{fig:mopa_znse}(d), the system maintains a 16-fold gain even with higher seed inputs, showing unexhausted gain-extraction capacity. For weak seeds (30~mW), the amplifier achieves watt-level output with a gain factor of 38. These results confirm Cr:ZnSe as a favorable choice for high-power mid-IR amplification, partly attributed to its lower quantum defect when pumped with a 1.94-{\textmu}m Thulium-doped fiber laser. This superiority is further reinforced by its higher optical quantum efficiency, the slower onset of thermally activated nonradiative decay with temperature, and a slightly larger emission cross-section compared to ZnS \cite{sorokina2004}, which together enable efficient and nearly linear power scaling.

\subsection{Dual-Stage MOPA System Based on Cr:ZnS/Cr:ZnSe Polycrystal Rods}

\begin{figure}[H]
    \centering
    \includegraphics[width=0.95\linewidth]{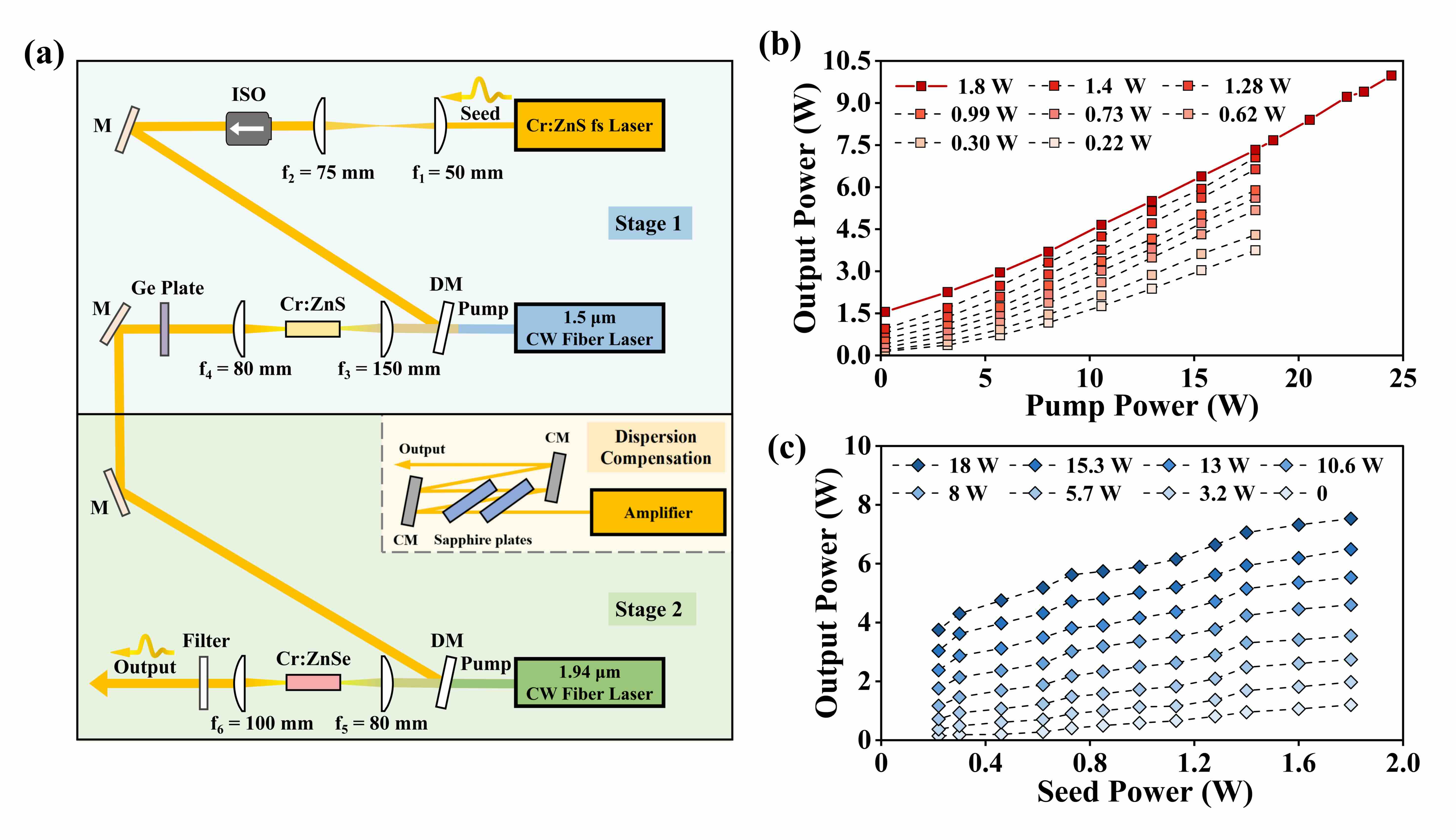}
    \caption{(a) Schematic diagram of the dual-stage Cr:ZnS/Cr:ZnSe MOPA system. 
    (b) Average output power of the second stage versus CW pump power at different seed powers from the first stage. 
    (c) Average output power of the second stage versus seed power from the first stage under different pump conditions.}
    \label{fig:dual_mopa}
\end{figure}
\begin{figure}[H]
    \centering
    \includegraphics[width=0.95\linewidth]{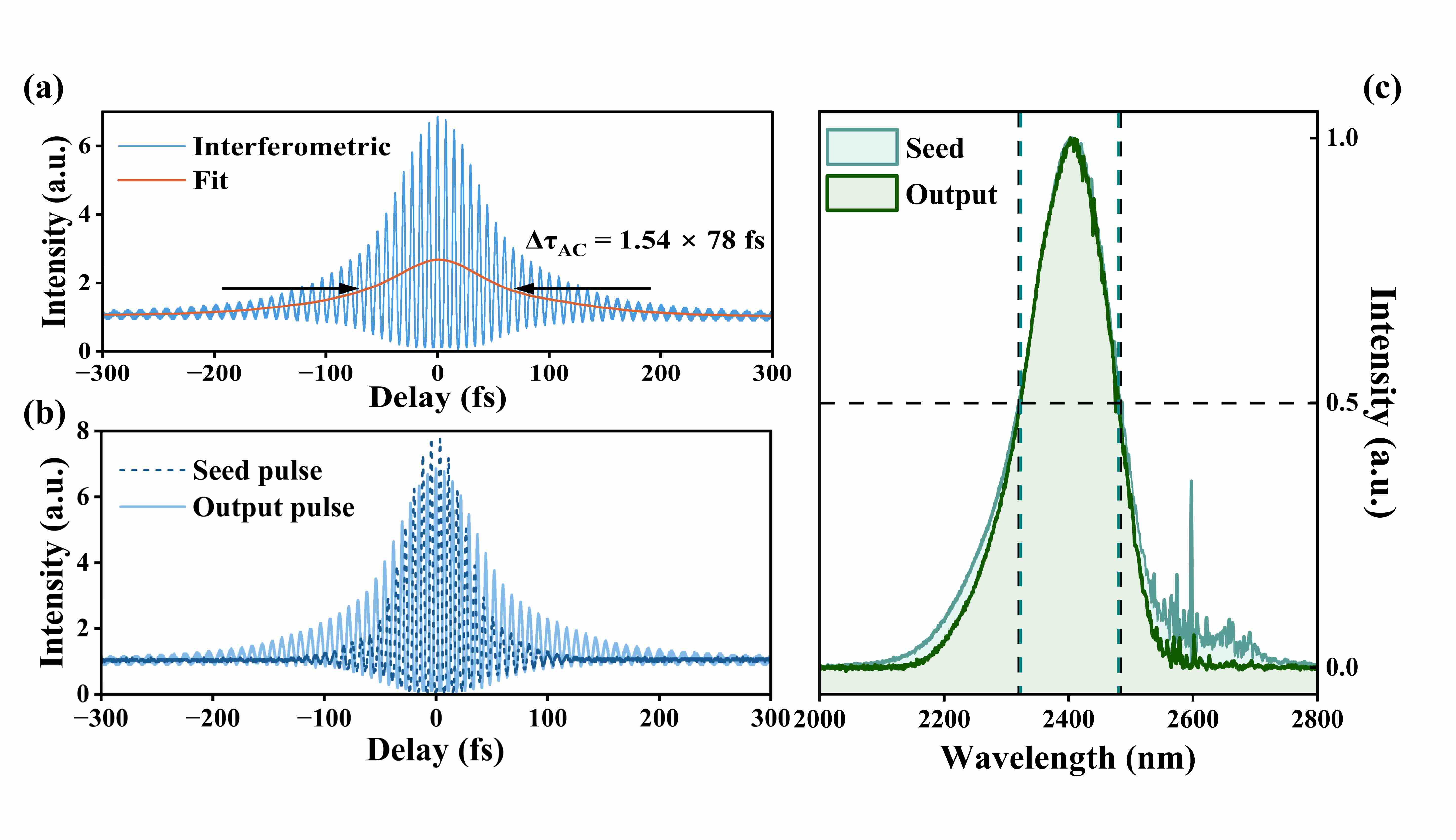}
    \caption{Temporal and spectral characterization of the dual-stage MOPA system: 
    (a) Interferometric autocorrelation of amplified pulses; 
    (b) Comparison of autocorrelation traces of seed pulses from the oscillator and amplified pulses after the second stage; 
    (c) Spectra of seed pulses from the oscillator and amplified pulses after the second stage.}
    \label{fig:dual_temporal}
\end{figure}
\begin{figure}[H]
    \centering
    \includegraphics[width=0.95\linewidth]{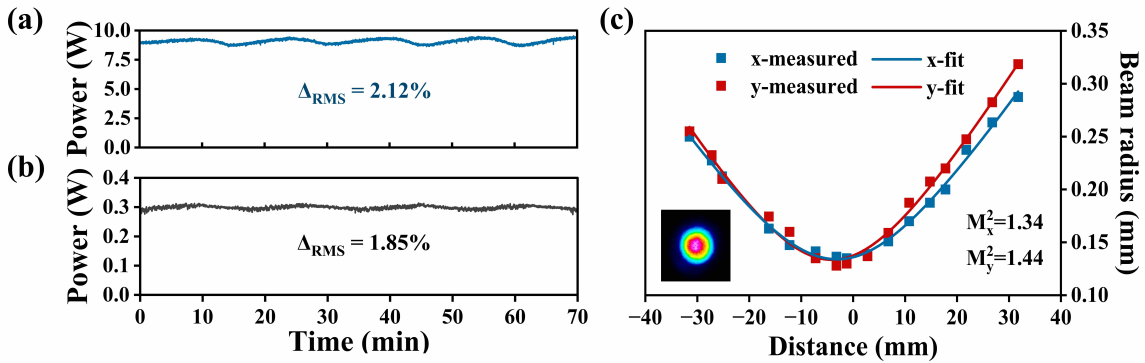}
    \caption{(a) Power stability of amplified dual-stage output. 
    (b) Power stability of the seed pulses. 
    (c) Beam quality characterization of amplified pulses.}
    \label{fig:dual_stability}
\end{figure}

For further power scaling, a two-stage MOPA architecture, cascading the previously discussed Cr:ZnS (Stage~1) and Cr:ZnSe (Stage~2) amplifier, along with a dispersion compensation module positioned post-Stage~2, is built, as shown in Fig.~\ref{fig:dual_mopa}(a). The femtosecond seed originates from the Cr:ZnS fs laser, first collimated and reshaped by a telescope formed by lenses ($f_1 = 50$~mm and $f_2 = 75$~mm), then unidirectionally transmitted via an optical isolator and redirected by plane gold mirrors into Stage~1. Here, the seed is co-propagated with a 1.55~{\textmu}m CW fiber laser pump, combined via a dichroic mirror and focused into the Cr:ZnS crystal by lenses ($f_3 = 150$~mm and $f_4 = 80$~mm) for initial gain extraction. After filtering out the residual 1.55~{\textmu}m pump signal using a thin Ge plate, the pre-amplified pulses then enter Stage~2, where they are merged with a 1.94~{\textmu}m CW fiber laser pump via another DM and focused into the Cr:ZnSe crystal by lenses ($f_5 = 80$~mm and $f_6 = 100$~mm) for secondary amplification. After removing the residual 1.94~{\textmu}m pump beam with a longpass filter, the output from Stage~2 passes through a dispersion compensation stage (see inset of Fig.\ref{fig:dual_mopa}(a)) before being extracted as the final output, ensuring the offset of accumulated positive dispersion in the polycrystal rods and thereby preventing pulse broadening and degradation of peak power and coherence.

As seen in Fig.~\ref{fig:dual_mopa}(b), the dual-stage MOPA system achieves a breakthrough in power scaling. The first stage (Stage 1) serves as a pre-amplifier (ranging from 0.22~W to 1.8~W) for the main amplifier (Stage 2). For all initial seed power levels, the output power increases nearly linearly with pump power up to 24~W. Higher seed powers result in higher output powers under identical pump conditions, as stronger seeds optimize pump extraction efficiency. Under 24~W pump power and 1.8~W seed input, the system delivers 10.4~W output, nearly 2.5 times the 4.2~W output of the standalone Cr:ZnSe system (see Fig.~\ref{fig:mopa_znse}(b)]), corresponding to a superior gain of 32.5$\times$ and an optical-to-optical efficiency of 43.3\%. Importantly, the absence of saturation at 1.8~W seed (Fig.~\ref{fig:dual_mopa}(b)) and 24~W pump confirms that neither the Cr:ZnS pre-amplifier nor the Cr:ZnSe main amplifier has reached its physical limits.

The dispersion compensation module (inset of Fig.~\ref{fig:dual_mopa}(a)) consists of two 5-mm-thick sapphire plates tilted at Brewster' angle and two chirped mirrors (CMs) (GDD=-250 fs\textsuperscript{2}). The combination of CMs and sapphire plates provide bulk negative group-delay dispersion (GDD), counteracting the accumulated positive dispersion during dual-stage amplification, while the chirped mirrors address residual and higher-order dispersion.

The autocorrelation trace of the amplified pulse after dispersion compensation indicates a pulse duration of 78~fs (see Fig.~\ref{fig:dual_temporal}(a)). By contrast, Fig.~\ref{fig:dual_temporal}(b) compares the seed pulse (39~fs) with the amplified pulse. The extended sidebands in the amplified pulse may be due to the uncompensated higher-order dispersions, mainly the accumulated third-order dispersion in Cr:ZnS/Cr:ZnSe and the sapphire plate used for the compensation of the second-order. While second-order GVD is compensated, residual higher-order phase distortions shift spectral components temporally, forming the observed sidebands. Fig.~\ref{fig:dual_temporal}(c) shows the spectra of seed and amplified pulses: the seed has a 162-nm FWHM, while the amplified output retains 158~nm (spectral narrowing $< 2.5$\%), confirming preserved spectral integrity during power scaling.

Finally, the measured output power stability of the dual-stage system is shown Fig.~\ref{fig:dual_stability}(a), with an RMS fluctuation of 183~mW at a mean power of 9.09~W indication a variance of 0.03, and a RMS instability of 2.12\%. For comparison, the seed (Fig.~\ref{fig:dual_stability}(b)) shows an RMS fluctuation of $5.5 \times 10^{-3}$~W, a mean power of 298~mW, a variance of $3.2 \times 10^{-5}$, and an RMS instability of 1.85\%. The periodic fluctuation of power on a $\sim$20 min scale is likely linked to the thermal management in the Erbium-doped fiber laser used as the pump source for the oscillator. These results demonstrate that the dual-stage amplifier preserves the power stability of the seed without introducing significant additional fluctuations. The characterization of the beam quality of the amplified fs pulses is shown in Fig.~\ref{fig:dual_stability}(c).

\section{Conclusion}

This study presents a dual-stage master oscillator power amplifier (MOPA) system employing Cr:ZnS/ZnSe crystal rods with engineered doping profiles, achieving average output powers over 10~W. Using 27-mm-long Cr:ZnS and 28-mm-long Cr:ZnSe polycrystals, amplification performances with three amplifier architectures of a Kerr-lens mode-locked master oscillator (39~fs pulses, 350~mW, 2~nJ pulse energy) are investigated. The dual-stage MOPA delivered 10.4~W output power, a 30-fold amplification of seed pulses, at 60.1~nJ pulse energies and repetition rates of 173-MHz with minimal gain narrowing. After compensating the second order dispersion, the output pulse duration was measured to be 78~fs. Given the ability to tailor the concentration profile, these engineered Cr:ZnS/ZnSe crystals shall pave ways for further power scaling of mid-IR ultrafast lasers and open new horizons for applications with high-power fs laser systems in the 2--3~$\mu$m wavelength range.

\section{Acknowledgement}

The authors wish to acknowledge the support of the Strategic Priority Research Program of the Chinese Academy of Sciences (XDB0650000), the National Key R\&D Program of China (2022YFB3605704, 2021YFB3501700), the Shanghai Pujiang Program (22PJ1414900). 

\newpage

\bibliography{references}

\end{document}